\documentclass[prd,aps,onecolumn,superscriptaddress,showpacs,dvips]{revtex4}
\usepackage{feynmp}
\usepackage{amssymb}
\usepackage{epsfig}
\usepackage{graphicx}
\usepackage{subfigure}
\usepackage{hyperref}
\usepackage{CJK}

\begin{document}
\begin{CJK*}{GBK}{song}

\begin{flushright}

USTC-ICTS-12-10\\

\end{flushright}

\title{Gravitational Corrections to $\Phi^{4}$ Theory
with Spontaneously Broken Symmetry}

\author{Hao-Ran Chang}
\email{hrchang@mail.ustc.edu.cn}
\affiliation{Department of Modern Physics,\\
University of Science and Technology of China,
 Hefei, Anhui, 230026, People's Republic of China}

\author{Wen-Tao Hou}
\affiliation{Department of Modern Physics,\\
University of Science and Technology of China,
 Hefei, Anhui, 230026, People's Republic of China}

\author{Yi Sun}
\affiliation{Department of Modern Physics,\\
University of Science and Technology of China,
 Hefei, Anhui, 230026, People's Republic of China}
\affiliation{Interdisciplinary Center for Theoretical Study,\\
University of Science and Technology of China, Hefei, Anhui,
230026, People's Republic of China}

%\date{\today}

\begin{abstract}
We consider a complex scalar $\Phi^4 $ theory with spontaneously
broken global U(1) symmetry, minimally coupling to perturbatively quantized Einstein gravity which is treated as an effective theory
at the energy well below the Planck scale. Both the lowest order
pure real scalar correction and the gravitational correction to the renormalization of the Higgs sector in this model have been investigated.
Our results show that the gravitational correction renders the renormalization of the Higgs sector in this model inconsistent while
the pure real scalar correction to it leads to a compatible renormalization.
\end{abstract}


%\keywords{gravitational correction; symmetry spontaneously breaking;
%renormalization.}

\maketitle

%%%%%%%%%%%%%%%%%%%%%%%%%%%%%Main Body%%%%%%%%%%%%%%%%%%%%%%%%%%%%%%%

\section{\label{introduction}Introduction}
It is well known that Einstein's general relativity quantized on
a fixed background is not a renormalizable quantum field theory
since the mass dimension of its coupling constant $\kappa=\sqrt{
32\pi G_{N}}$ is negative \cite{DeWittDuffChrist, tHooftVeltman-1}.
Furthermore, the coupling of the Einstein-Hilbert theory to any
type of matter fields leads to nonrenormalizable theories as well
\cite{tHooftVeltman-1,Deser1,Deser2}. Provided that our interest
is restricted to the physics at the energy well below the Planck
scale $M_{Planck}={G_{N}^{-1/2}}\approx10^{19}\mathrm{GeV}$,
perturbatively quantized gravity can be treated as an effective
field theory, which has been established by Donoghue \cite{Donoghue}.
It can provide the results that should coincide with the results predicted by the underlying fundamental quantum theory of gravitation.

In their initiative paper \cite{RobinsonWilczek}, within the
framework of the effective quantum theory of gravity, Robinson and
Wilczek presented a calculation that claimed the behavior of
running gauge coupling constants in Abelian and non-Abelian gauge theories could be altered by quantum gravitational correction
which would render all gauge theories asymptotically free. However, doubts have been cast on their conclusion by some authors and the
result has been studied from different approaches. After careful reconsideration of the calculation, Pietrykowski \cite{Pietrykowski} first showed that Robinson and Wilczek's result was not gauge
condition independent, and that the gravitational correction to
the $\beta$ function at one-loop order was absent in the harmonic
gauge. Using a gauge condition independent background field method \cite{VilkoviskyDeWitt}, along with dimensional regularization (DR) \cite{tHooftVeltman-2}, Toms \cite{Toms2007} showed that it did
not lead to nonvanishing gravitational contributions to the
running of gauge coupling constants. This result has been confirmed
by a traditional Feynman diagram approach calculation using standard
Feynman rules in Ref.\cite{Ebert2008}, where if a momentum space
cutoff was used the quadratic divergences could be made to cancel,
leaving a result that was consistent with DR. Subsequent works of
Toms \emph{et al}. have investigated the cases with the cosmological
constant \cite{Toms2008,MackayToms}. Further, various approaches were
used to discuss the applications to the gravitational corrections
to a series of theories \cite{WuFeng, Ebert2009, RodigastSchuster,
Toms2010, BhattPatraSarkar, ShaposhnikovWetterich,ZZVP,DaumHarstReuter,
ADE, Gerwick2010, EM2010, Folkerts2011, FBMN, CalmetYang, TangWu,
He2010,Toms2011}.

In Ref.\cite{ADE}, the authors concluded that outside of some
special cases, such as the ordinary $\varphi^4$ interaction,
the gravitational contribution to the running coupling constant
of other theories, such as Yukawa and gauge theories, is not a
useful and universal idea in the perturbative regime. So special attention should be paid to the gravitational corrections to
the $\varphi^4$ interaction. Moreover, Rodigast and Schuster \cite{RodigastSchuster} considered gravitational corrections to
the ordinary $\varphi^4$ interaction and related the real scalar
field to the Higgs boson (hereafter Higgs field is denoted by H).
However, since there is not only H$^{4}$ interaction but also
H$^{3}$ interaction in the standard model (SM) because of the
spontaneous breaking of electroweak SU(2)$_{L}\times$U(1)$_{Y}$
symmetry, a relatively physical SM-like case for the Higgs sector is
a model in which the three real scalars interaction as well as the
four real scalars interaction should be included. Motivated by
this, following the approach of Ref.\cite{RodigastSchuster},
we investigate $\Phi^4 $ theory with spontaneously broken global
U(1) symmetry, minimally coupling to perturbatively quantized
gravity which is treated as an effective theory. We study the
lowest order pure real scalar correction and gravitational correction
to the renormalization of the Higgs sector in this model. It has been
found that the gravitational correction renders the renormalization
of this model inconsistent while the pure real scalar correction to it
leads to a consistent renormalization.

This paper is organized as follows. First, we provide the
framework of calculating the quantum corrections to our model
in Sec.~\ref{framework}. Then in Sec.~\ref{corrections}, the
pure real scalar correction and the gravitational correction to this
model are studied. In Sec.~\ref{comparisonanddiscussion}, the
comparison with scalar quantum electrodynamics (SQED) with
spontaneously broken global U(1) symmetry and discussion are
presented. Finally, we give our conclusion in Sec.~\ref{conclusion}.

\section{\label{framework}Framework of calculation}
Before investigating our model, we first sketch the
approach of Ref.\cite{RodigastSchuster}. The full original
Lagrangian takes the following form:
\begin{eqnarray}
\mathcal{L}&=&
\mathcal{L}_{EH}+\mathcal{L}_{gf}+\mathcal{L}_{ghost}
+\mathcal{L}_{gs}.\label{Equ:fulllagrangian}
\end{eqnarray}
In Eq.(\ref{Equ:fulllagrangian}), the Einstein-Hilbert
Lagrangian $\mathcal{L}_{EH}$ reads
\begin{eqnarray}
\mathcal{L}_{EH}&=&\frac{2}{\kappa^2}\sqrt{-g}~R,
\end{eqnarray}
where $g$ is the determinant of the metric $g_{\mu\nu}$
and $R=g^{\mu\nu}\mathcal{R}_{\mu\nu}$ is the Ricci scalar
defined by
\begin{eqnarray}
\mathcal{R}_{\mu\nu}&=&
\partial_{\mu}\Gamma_{\rho\nu}^{\rho}
-\partial_{\rho}\Gamma_{\mu\nu}^{\rho}
+\Gamma_{\mu\lambda}^{\rho}\Gamma_{\rho\nu}^{\lambda}
-\Gamma_{\rho\lambda}^{\rho}\Gamma_{\mu\nu}^{\lambda},\nonumber \\
\Gamma_{\mu\nu}^{\rho}&=&\frac{1}{2}g^{\rho\sigma}
(\partial_{\mu}g_{\nu\sigma}+\partial_{\nu}g_{\mu\sigma}
-\partial_{\sigma}g_{\mu\nu}).\nonumber
\end{eqnarray}
In order to quantize gravity, the metric $g_{\mu\nu}$ should
be perturbed about the flat Minkowski background $\eta_{\mu\nu}$,
then
\begin{eqnarray}
g_{\mu\nu}&=&\eta_{\mu\nu}+\kappa h_{\mu\nu},\label{Equ:expasion}
\end{eqnarray}
with the symmetric tensor field $h_{\mu\nu}$ being the graviton,
the spacetime fluctuations. Note that hereafter indices are
raised and lowered with the background metric $\eta_{\mu\nu}$.
The inverse metric becomes
\begin{eqnarray}
g^{\mu\nu}&=&\eta^{\mu\nu}-\kappa h^{\mu\nu}+\kappa^{2}
h^{\mu\alpha}h_{\alpha}^{\nu}+\mathcal{O}(\kappa^{3}),
\label{Equ:metricup}
\end{eqnarray}
and to $\mathcal{O}(\kappa^{2})$ the expansion of the measure
in terms of $h_{\mu\nu}$ is given by
\begin{eqnarray}
\sqrt{-g}&=&1+\frac{1}{2}\kappa\eta^{\mu\nu}h_{\mu\nu}
-\frac{1}{2}h_{\alpha\beta} \frac{\kappa^{2}}{2}P^{\alpha\beta\mu\nu}h_{\mu\nu},
\label{Equ:measure}
\end{eqnarray}
with
\begin{eqnarray}
P^{\alpha\beta\mu\nu}&=&
\frac{1}{2}\left(\eta^{\alpha\mu}\eta^{\beta\nu}
+\eta^{\alpha\nu}\eta^{\beta\mu}
-\eta^{\alpha\beta}\eta^{\mu\nu}\right).
\end{eqnarray}
General coordinate invariance implies that $\mathcal{L}_{EH}$
is invariant under the infinitesimal transformation
\begin{eqnarray}
\delta_{\xi}h_{\mu\nu}&=&
h_{\sigma\mu}\partial_{\nu}\xi^{\sigma}
+h_{\sigma\nu}\partial_{\mu}\xi^{\sigma}
+\xi^{\sigma}\partial_{\sigma}h_{\mu\nu}\nonumber\\
&&+\frac{1}{\kappa}(\partial_{\mu}\xi_{\nu}
+\partial_{\nu}\xi_{\mu}).
\end{eqnarray}
The Faddeev-Popov procedure \cite{FaddeevPopov} is used to fix
this gauge freedom by employing the harmonic (de Donder) gauge
fixing condition
\begin{eqnarray}
G_{\mu}&=&\partial^{\nu} h_{\mu\nu}
-\frac{1}{2}\partial_{\mu} h^{\nu}_{\nu},
\end{eqnarray}
and this leads to the gauge fixing term Lagrangian
$\mathcal{L}_{gf}$ as well as the corresponding ghost
term Lagrangian $\mathcal{L}_{ghost}$,
\begin{eqnarray}
\mathcal{L}_{gf}&=&G_{\mu}G^{\mu},\nonumber\\
\mathcal{L}_{ghost}&=&-\bar{c}^{\mu}
\left(\frac{\delta G_{\mu}}{\delta \xi^{\nu}}\right)c^{\nu}.
\end{eqnarray}

\begin{figure}[htbp]
    \centering
    \includegraphics[width=3.5cm]{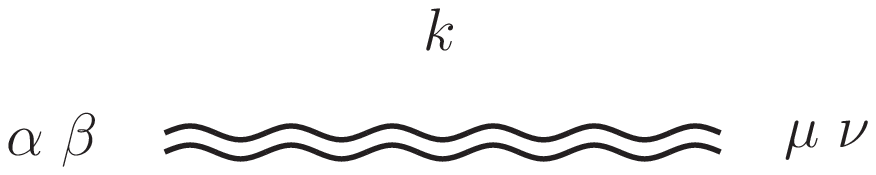}
    \caption{Graviton propagator}
    \label{Fig:propagator}
\end{figure}

Then in the harmonic gauge the graviton propagator shown
in Fig.(\ref{Fig:propagator}) takes the form
\begin{eqnarray}
D^{\alpha\beta,\mu\nu}(k)&=&
\frac{i}{k^{2}}~P^{\alpha\beta\mu\nu}.
\end{eqnarray}

The graviton-scalar Lagrangian is
\begin{eqnarray}
\mathcal{L}_{gs}=\sqrt{-g}\Bigl[
\frac{1}{2}g^{\mu\nu}(\partial_{\mu}\varphi)(\partial_{\nu}\varphi)
-\frac{1}{2}m^{2}\varphi^{2} -\frac{\lambda}{4!}\varphi^{4}\Bigl].
\label{Equ:varphilagrangian}
\end{eqnarray}

\begin{figure}[htbp]
   \centering
   \raisebox{-2.75ex}{\includegraphics[width=8cm]{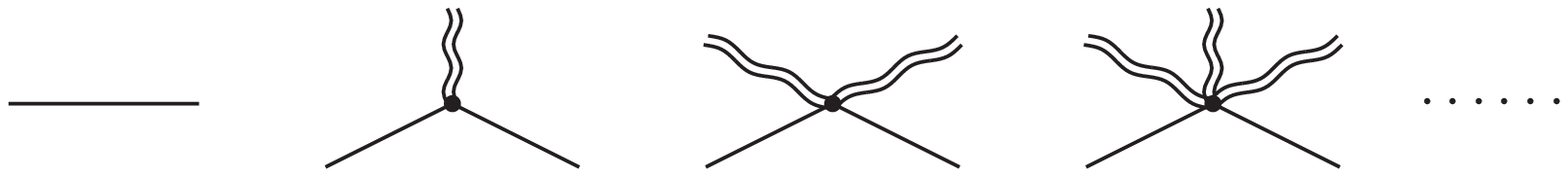}}\\
   \hspace{0.5cm}(a)\hspace{1.2cm}(b)\hspace{1.5cm}
   (c)\hspace{1.5cm}(d)\hspace{1.5cm}
   \caption{Two scalars can couple to any number of gravitons.
   (a)-(d) are of order $\mathcal{O}(\kappa^{0})$, $\mathcal{O}(\kappa^{1})$, $\mathcal{O}(\kappa^{2})$ and $\mathcal{O}(\kappa^{3})$, respectively. Hereafter the solid
   line represents Higgs field $\phi$ and the double wavy line
   denotes graviton $h_{\mu\nu}$.}\label{Fig:kappas}
\end{figure}

Expanding Eq.(\ref{Equ:varphilagrangian}) in orders of $\kappa$
leads to an infinite series of interactions involving arbitrary
numbers of gravitons, e.\,g., as shown in Fig.(\ref{Fig:kappas}),
two scalars can couple to any number of gravitons.

In the calculation, the gravitational ghosts are irrelevant since
only one-loop diagrams with no external gravitons are concerned.
The remaining thing to do is to calculate the renormalizations of the scalar
field and coupling constant using the standard Feynman rules approach
in the minimal subtraction scheme and the DR scheme with spacetime
dimension $D=4-2\epsilon$, where the one-loop divergences manifest themselves as poles at $D=4$, in which the $Z$ factors contain
solely the divergent pole terms proportional to $\frac{1}{\epsilon}$.
One subtle issue is that because the squared-momentum-dependent terms
appear in the $\mathcal{O}(\kappa^{2})$ correction to the four-point
function of the $\varphi^{4}$ interaction there should be one
possible higher-derivative counterterm of the form
\begin{eqnarray}
\mathcal{L}_{hdc}\sim \sqrt{-g}~g^{\mu\nu}
~(\partial_{\mu}\varphi)(\partial_{\nu}\varphi)\varphi^{2},
\label{Equ:hdct}
\end{eqnarray}
to remove the squared-momentum-dependent terms appearing in the $\mathcal{O}(\kappa^{2})$ correction \cite{schuster2008dpl}.

Different from that in Ref.\cite{RodigastSchuster}, we consider
a complex scalar $\Phi^{4}$ theory with spontaneously broken
global U(1) symmetry. The graviton-scalar Lagrangian for it
is taken to be
\begin{eqnarray}
\mathcal{L}_{gs}&=&\sqrt{-g}\Bigl[g^{\mu\nu}
(\partial_{\mu}\Phi)^{*}(\partial_{\nu}\Phi)-V(\Phi)\Bigl],
\label{Equ:Philagrangian}
\end{eqnarray}
with the potential taking the form
\begin{eqnarray}
V(\Phi)&=&\frac{\lambda}{2}(\Phi^{*}\Phi-\frac{v^{2}}{2})^{2},
\label{Equ:potential}
\end{eqnarray}
where
\begin{eqnarray}
v&=&\sqrt{\frac{2\mu^{2}}{\lambda}}.
\end{eqnarray}

Evidently, this Lagrangian is invariant under the global
U(1) transformation
\begin{eqnarray}
\Phi\to \Phi^{'}=e^{-i\theta}\Phi,
\label{Equ:transformation}
\end{eqnarray}
where $\theta$ is independent of spacetime coordinate $x$.

It is straightforward to obtain the vacuum expect value(VEV)
of $\Phi$ at tree-level,
\begin{eqnarray}
<0|\Phi|0>&=&\pm\frac{v}{\sqrt{2}}.
\label{Equ:VEV}
\end{eqnarray}

We can express $\Phi$ in terms of the real fields $\Phi_{1}$
and $\Phi_{2}$
\begin{eqnarray}
\Phi&=&\frac{\Phi_{1}+i\Phi_{2}}{\sqrt{2}},
\label{Equ:decomposion}
\end{eqnarray}
and then choose the vacuum,
\begin{eqnarray}
<0|\Phi_{1}|0>&=&v,\nonumber\\
<0|\Phi_{2}|0>&=&0.
\label{Equ:vacuum}
\end{eqnarray}

By paralleling the procedure carried out in
Ref.\cite{BarinPassarino}, after the spontaneous
breaking of global U(1) symmetry, in the Lagrangian,
a tadpole constant $\delta_{t}$ should appear that
is zero in the lowest order and must be adjusted in
such a way that the VEV of the Higgs field remains
zero order by order in perturbation theory, then
\begin{eqnarray}
\Phi_{1}&=&\phi+v(1+\delta_{t}),\nonumber\\
\Phi_{2}&=&\rho.\label{Equ:barefield}
\end{eqnarray}
To $\mathcal{O}(\delta)$, Eq.(\ref{Equ:Philagrangian})
can be reexpressed in terms of the Higgs field $\phi$
and Goldstone field $\rho$,
\begin{eqnarray}
\mathcal{L}_{gs}&=&\sqrt{-g}\Bigl[
\frac{1}{2}g^{\mu\nu}(\partial_{\mu}\phi)(\partial_{\nu}\phi)
+\frac{1}{2}g^{\mu\nu}(\partial_{\mu}\rho)(\partial_{\nu}\rho)
\nonumber\\&&
-\frac{\lambda}{8}\phi^{4}-\frac{\lambda}{2}v^{2}\phi^{2}
-\frac{\lambda}{8}\rho^{4}-\frac{\lambda}{2}v\phi^{3}
-\frac{\lambda}{4}\phi^{2}\rho^{2}\nonumber\\&&
-\frac{\lambda}{2}v\phi\rho^{2}
-\lambda v^{2}\phi^{2}\delta_{t}
-\frac{\lambda}{2}v\phi^{3}\delta_{t}
-\frac{\lambda}{2}v^{2}\phi^{2}\delta_{t}\nonumber\\&&
-\lambda v^{3}\phi\delta_{t}
-\frac{\lambda}{2}v\phi\rho^{2}\delta_{t}
-\frac{\lambda}{2}v^{2}\rho^{2}\delta_{t}\Bigl].
\label{Equ:SSBlagrangian}
\end{eqnarray}

Obviously, there is no cosmological term in this model,
so the expansion of the metric around a flat Minkowski
background, as done in Eq.(\ref{Equ:expasion}), is valid.
In order to investigate the contribution of the gravitational
correction to the renormalization of the Higgs sector in
$\Phi^{4}$ theory with spontaneously broken global U(1)
symmetry, we replace the quantities $\phi$, $\rho$,
$\lambda$, and $v$ by the corresponding bare quantities
$\phi_{0}$, $\rho_{0}$, $\lambda_{0}$, and $v_{0}$,
respectively, but leave the graviton field $h_{\mu\nu}$
and the gravitational coupling $\kappa$ unchanged as
done in Ref.\cite{RodigastSchuster}. This treatment is
reasonable since their contributions of gravitational
corrections to the renormalization of the quantities
$h_{\mu\nu}$ and $\kappa$ are higher order and therefore
can be neglected.

It is convenient to introduce the following relations
\begin{eqnarray}
\phi_{0}&=&\sqrt{Z_{\phi}}~\phi,\nonumber\\
\rho_{0}&=&\sqrt{Z_{\rho}}~\rho,\nonumber\\
\lambda_{0}&=&Z_{\phi}^{-2}\lambda Z_{\lambda},\nonumber\\
v_{0}&=&\sqrt{Z_{\phi}}~v Z_{v},
\label{Equ:renormalizations}
\end{eqnarray}
with
\begin{eqnarray}
Z_{\phi}&=&1+\delta_{\phi},\nonumber\\
Z_{\rho}&=&1+\delta_{\rho},\nonumber\\
Z_{\lambda}&=&1+\delta_{\lambda},\nonumber\\
Z_{v}&=&1+\delta_{v}.\label{Equ:relations}
\end{eqnarray}

The $Z$ factors (or equivalently $\delta$ factors) and
$\delta_{t}$ are used to absorb all the ultraviolet
divergences arising from the pure real scalar correction
and gravitational correction in the model. Since we are interested
only in one-loop diagrams with no external
gravitons and the lowest order gravitational corrections
to the renormalizations of $\phi$, $\lambda$, and $v$,
other terms with both $\delta$'s and $h_{\mu\nu}$ or
without $\phi$, $\rho$ in Eq.(\ref{Equ:SSBlagrangian})
are omitted, hence to $\mathcal{O}(\delta)$ in terms of
renormalized quantities, the part of Eq.(\ref{Equ:SSBlagrangian}) relevant for our considerations reads explicitly
\begin{eqnarray}
\mathcal{L}_{gs}&=&\sqrt{-g}\Bigl[
\frac{1}{2}g^{\mu\nu}(\partial_{\mu}\phi)(\partial_{\nu}\phi)
+\frac{1}{2}g^{\mu\nu}(\partial_{\mu}\rho)(\partial_{\nu}\rho)
-\frac{\lambda}{8}\phi^{4}\nonumber\\&&
-\frac{\lambda}{2}v^{2}\phi^{2}
-\frac{\lambda}{8}\rho^{4}
-\frac{\lambda}{2}v\phi^{3}
-\frac{\lambda}{4}\phi^{2}\rho^{2}
-\frac{\lambda}{2} v\phi\rho^{2}\Bigl]\nonumber\\&&
-\lambda v^{3}\phi\delta_{t}
+\frac{1}{2}(\partial_{\mu}\phi)^{2}\delta_{\phi}
-\frac{\lambda}{2} v^{2}\phi^{2}
(\delta_{\lambda}+2\delta_{v}+3\delta_{t})\nonumber\\&&
-\frac{\lambda}{2} v\phi^{3}
(\delta_{\lambda}+\delta_{v}+\delta_{t})
-\frac{\lambda}{8} \phi^{4}\delta_{\lambda}
+\frac{1}{2}(\partial_{\mu}\rho)^{2}\delta_{\rho}\nonumber\\&&
-\frac{\lambda}{2} v^{2}\rho^{2}\delta_{t}
-\frac{\lambda}{4} \phi^{2}\rho^{2}
(\delta_{\lambda}-\delta_{\phi}+\delta_{\rho})\nonumber\\&&
-\frac{\lambda}{2} v\phi\rho^{2}
(\delta_{\lambda}+\delta_{v}-\delta_{\phi}
+\delta_{\rho}+\delta_{t})\nonumber\\&&
-\frac{\lambda}{8} \rho^{4}
(\delta_{\lambda}-2\delta_{\phi}+2\delta_{\rho})+\cdots,
\label{Equ:lagrangianfinal}
\end{eqnarray}
where the ellipsis represents the terms with both $\delta$'s
and $h_{\mu\nu}$, the constant terms without $\phi$, $\rho$
or $h_{\mu\nu}$, and the higher order terms involving scalars
as well (e.g., higher-derivative counterterms).

\begin{figure}[htbp]
    \centering
    \hspace{0.5cm}
    \subfigure{\includegraphics[width=1.25cm]{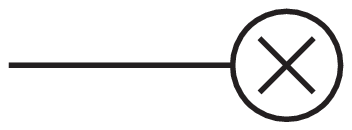}}
    \hspace{0.75cm}
    \subfigure{\includegraphics[width=2.5cm]{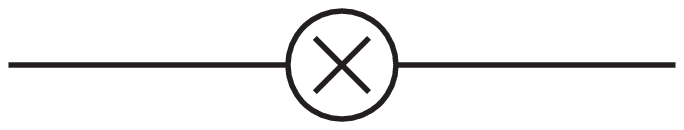}}\\
    (a)\hspace{2.4cm}(b)\\
    \subfigure{\includegraphics[width=2.5cm]{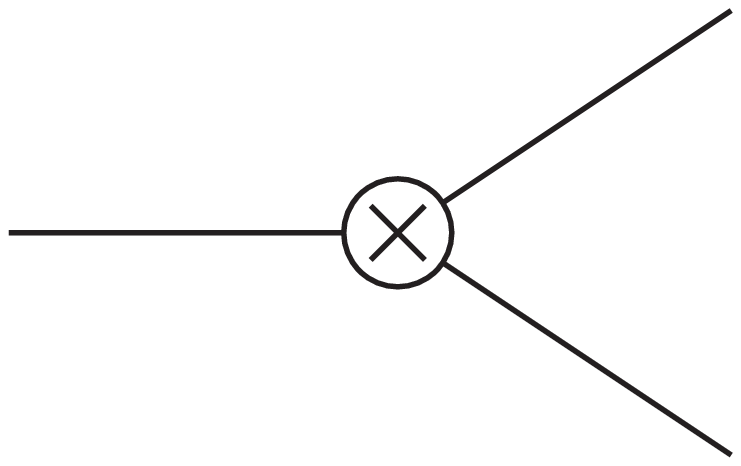}}
    \hspace{0cm}
    \subfigure{\includegraphics[width=2.4cm]{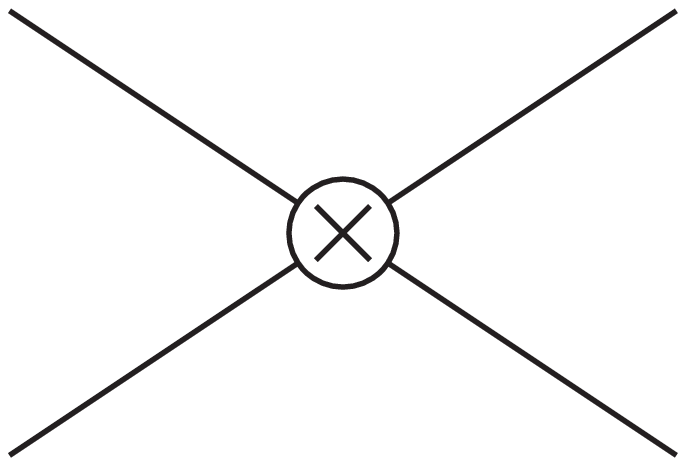}}\\
    (c)\hspace{2.4cm}(d)\\
    \caption{Counterterms for the one-point, two-point,
    three-point, and four-point functions of $\phi$.}
    \label{Fig:counterterms}
\end{figure}

From Eq.(\ref{Equ:lagrangianfinal}), the counterterms for
one-point, two-point, three-point, and four-point functions
of Higgs field $\phi$ listed in Fig.(\ref{Fig:counterterms})
are
\begin{eqnarray}
i\delta\Gamma^{(1)}&=&-i\lambda v^{3}\delta_{t},\nonumber\\
i\delta\Gamma^{(2)}&=&i[p^{2}\delta_{\phi}-\lambda v^{2}
(\delta_{\lambda}+2\delta_{v}+3\delta_{t})],\nonumber\\
i\delta\Gamma^{(3)}&=&-3i\lambda
v\left(\delta_{\lambda}+\delta_{v}+\delta_{t}\right),\nonumber\\
i\delta\Gamma^{(4)}&=&-3i\lambda\delta_{\lambda}.
\label{Equ:counterterms}
\end{eqnarray}

It should be pointed out that these four counterterms are valid
for both pure real scalar correction and gravitational correction.
And in our model there is not only $\phi^{4}$ but also $\phi^{3}$
coupling to any number of gravitons, which is one of the main
differences from Ref.\cite{RodigastSchuster}.

\section{\label{corrections}Pure real scalar correction VS
Gravitational correction}
Before considering the gravitational correction
part, we should investigate the pure real scalar case,
namely, the $\mathcal{O}(\kappa^{0})$ part, for comparison. Equation(\ref{Equ:lagrangianfinal}) reduces to the following
pure real scalar case:
\begin{eqnarray}
\mathcal{L}_{ps}&=&
\frac{1}{2}(\partial_{\mu}\phi)^{2}
+\frac{1}{2}(\partial_{\mu}\rho)^{2}
-\frac{\lambda}{8}\phi^{4}
-\frac{\lambda}{2}v^{2}\phi^{2}
-\frac{\lambda}{8}\rho^{4}\nonumber\\&&
-\frac{\lambda}{2}v\phi^{3}
-\frac{\lambda}{4}\phi^{2}\rho^{2}
-\frac{\lambda}{2} v\phi\rho^{2}
-\lambda v^{3}\phi\delta_{t}
+\frac{1}{2}(\partial_{\mu}\phi)^{2}\delta_{\phi}
\nonumber\\&&
-\frac{\lambda}{2} v^{2}\phi^{2}
(\delta_{\lambda}+2\delta_{v}+3\delta_{t})
-\frac{\lambda}{2} v\phi^{3}
(\delta_{\lambda}+\delta_{v}+\delta_{t})\nonumber\\&&
-\frac{\lambda}{8} \phi^{4}\delta_{\lambda}
+\frac{1}{2}(\partial_{\mu}\rho)^{2}\delta_{\rho}
-\frac{\lambda}{4} \phi^{2}\rho^{2}
(\delta_{\lambda}-\delta_{\phi}+\delta_{\rho})\nonumber\\&&
-\frac{\lambda}{2} v^{2}\rho^{2}\delta_{t}
-\frac{\lambda}{2} v\phi\rho^{2}
(\delta_{\lambda}+\delta_{v}-\delta_{\phi}
+\delta_{\varphi}+\delta_{t})\nonumber\\&&
-\frac{\lambda}{8} \rho^{4}
(\delta_{\lambda}-2\delta_{\phi}+2\delta_{\rho})+\cdots.
\label{Equ:Slagrangianfinal}
\end{eqnarray}

It is interesting to note that there is no odd number
of Goldstone field(s) in every term of the pure scalar
Lagrangian above, namely, the Goldstone field must appear
in pairs. Taking into account that groups O(2) and U(1) are locally
isomorphic while O(1) and Z$_{2}$ are exactly the same,
according to Goldstone theorem \cite{Goldstone}, after spontaneous symmetry breaking,
the original U(1) [or equivalently O(2)] symmetry is broken, but a residual
Z$_{2}$ [or equivalently O(1)] symmetry is still respected by the
Goldstone field. As a result, terms with not only even
but also odd numbers of Higgs fields can survive but terms
with odd numbers of Goldstone field(s) are forbidden.

\begin{figure}[htbp]
    \centering
    \subfigure{\includegraphics[width=1.75cm]{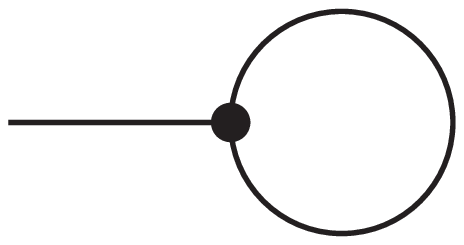}}
    \subfigure{\includegraphics[width=1.75cm]{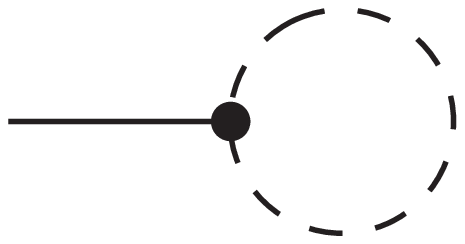}}\\
    \hspace{0.75cm}(a)\hspace{1.5cm}(b)\\
    \caption{$\mathcal{O}(\lambda)$ corrections to the one-point
    function of $\phi$. The diagrams (a) and (b) are both of
    weight $\frac{1}{2}$, which leads to a factor $\frac{1}{2}$
    to them. Hereafter the dashed line represents Goldstone field
    $\rho$.}
    \label{Fig:purescalaronepoint}
\end{figure}

\begin{figure}[htbp]
    \centering
    \subfigure{\includegraphics[width=2.5cm]{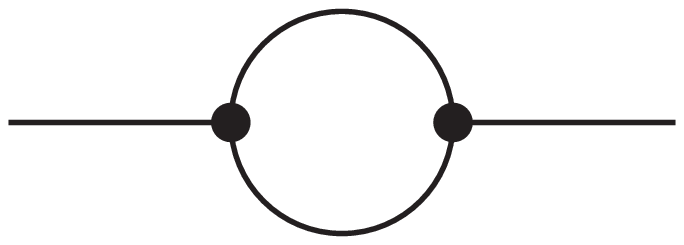}}
    \subfigure{\includegraphics[width=2.5cm]{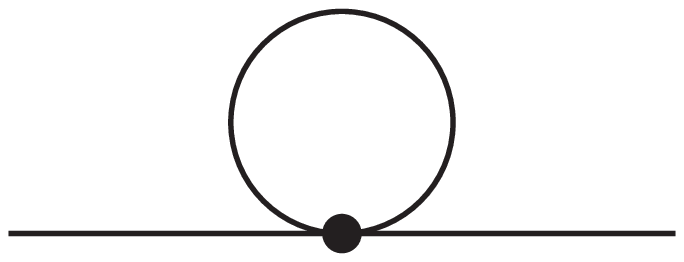}}\\
    (a)\hspace{2.3cm}(b)\\
    \subfigure{\includegraphics[width=2.5cm]{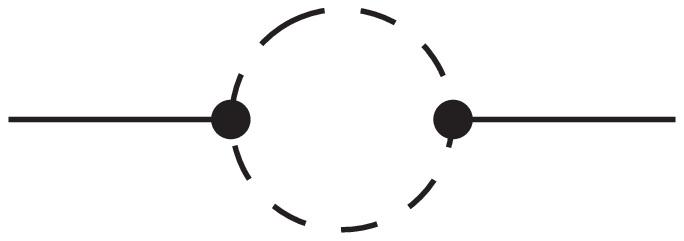}}
    \subfigure{\includegraphics[width=2.5cm]{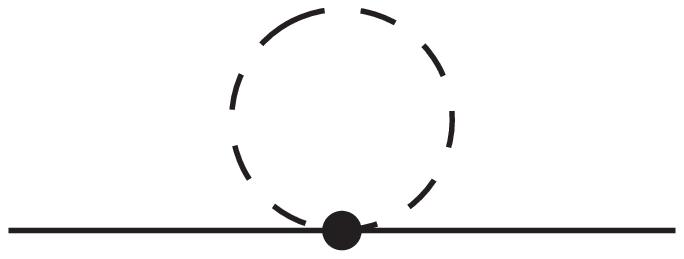}}\\
    (c)\hspace{2.3cm}(d)\\
    \caption{$\mathcal{O}(\lambda)$ corrections to the two-point
    function of $\phi$. (a)-(d) are all of weight $\frac{1}{2}$,
    which leads to a factor $\frac{1}{2}$.}
    \label{Fig:purescalartwopoint}
\end{figure}

\begin{figure}[htbp]
    \centering
    \subfigure{\includegraphics[width=2.5cm]{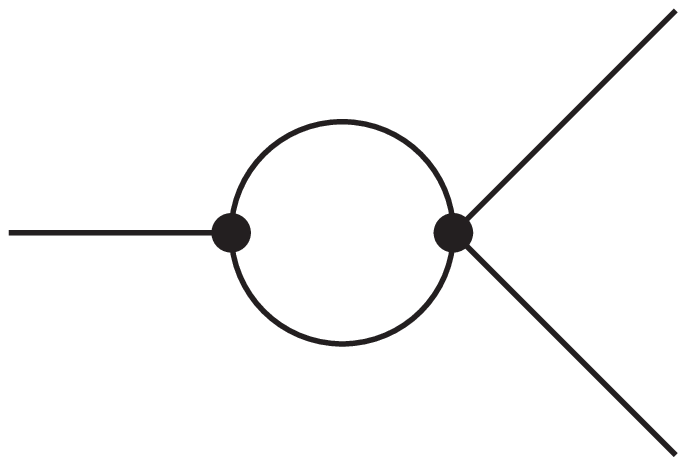}}
    \subfigure{\includegraphics[width=2.5cm]{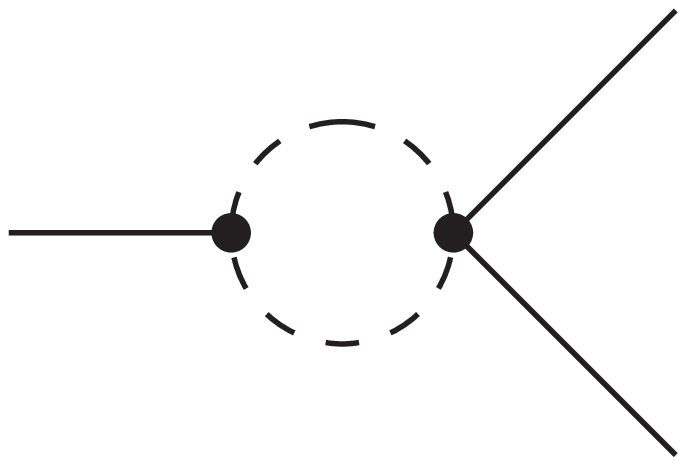}}\\
    (a)\hspace{2.3cm}(b)\\
    \caption{$\mathcal{O}(\lambda)$ correction to the three-point
    functions of $\phi$. (a) and (b) are both of
    weight $\frac{1}{2}$ and 3 permutations, which leads to a
    factor $\frac{3}{2}$.}
    \label{Fig:purescalarthreepoint}
\end{figure}

\begin{figure}[htbp]
    \centering
    \subfigure{\includegraphics[width=2.5cm]{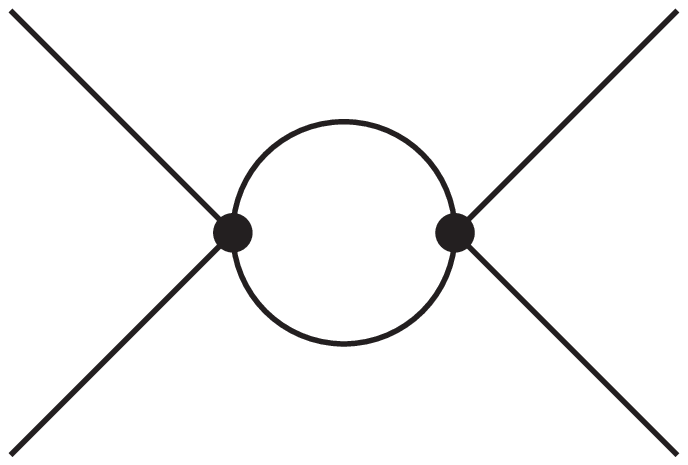}}
    \subfigure{\includegraphics[width=2.5cm]{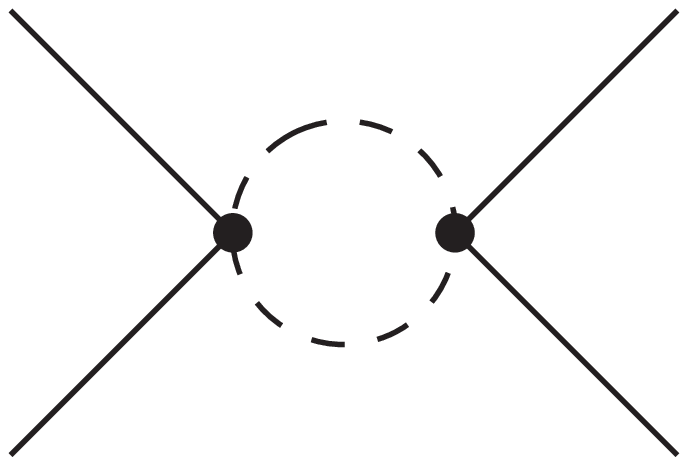}}\\
    (a)\hspace{2.3cm}(b)\\
    \caption{$\mathcal{O}(\lambda)$ corrections to four-point
    functions of $\phi$. The diagrams (a) and (b) are both of
    weight $\frac{1}{2}$ and 3 permutations, which leads to a
    factor $\frac{3}{2}$.}
    \label{Fig:purescalarfourpoint}
\end{figure}

The one-point, two-point, three-point and four-point
functions corresponding to the diagrams listed in Fig.(\ref{Fig:purescalaronepoint}),
Figs.\ref{Fig:purescalartwopoint}-\ref{Fig:purescalarfourpoint}) are in turn,
\begin{eqnarray}
i\Gamma_{s}^{(1)}&=&
\frac{3 i \lambda^{2} v^{2}}{2(4\pi)^{2}\epsilon},\nonumber\\
i\Gamma_{s}^{(2)}&=&
\frac{13 i \lambda^{2}v^{2}}{2(4\pi)^{2}\epsilon},\nonumber\\
i\Gamma_{s}^{(3)}&=&
\frac{10 i\lambda^{2}v}{2(4\pi)^{2}\epsilon},\nonumber\\
i\Gamma_{s}^{(4)}&=&
\frac{10 i\lambda^{2}}{2(4\pi)^{2}\epsilon}.
\label{Equ:scalargreenfunctions}
\end{eqnarray}

From one-point, two-point and three-point functions,
it is easy to get
\begin{eqnarray}
\delta_{t}&=&\frac{3 \lambda}{2(4\pi)^2\epsilon},\nonumber\\
\delta_{\phi}&=&0,\nonumber\\
\delta_{\lambda}&=&\frac{10 \lambda}{2(4\pi)^2\epsilon},\nonumber\\
\delta_{v}&=&-\frac{3 \lambda}{2(4\pi)^2\epsilon}.
\label{Equ:purescalarresult-3}
\end{eqnarray}

And exactly the same results are easy to reach from one-point, two-point, and four-point functions. To put it differently, the
pure real scalar correction to the renormalization of the Higgs sector
is consistent. The result is natural but nontrivial because both the coupling constants and the interactions of two-point, three-point
and four-point pure real scalar interactions after the spontaneous
breaking of global U(1) symmetry obey some strong constrains which
are dictated by the original symmetry. Even though the original
global U(1) symmetry is no longer apparent, it is still respected
in such a special way. However, if the coupling constants of
two-point, three-point, and four-point real scalar interactions are
corrected by an alien field which does not carry any information
of the original symmetry, one cannot expect it to render the
correction consistent.

Using the approach presented in Sec.~\ref{framework}, we consider
the lowest order gravitational correction to the renormalization of
the Higgs sector of $\Phi^4$ theory with spontaneously broken global
U(1) symmetry. Below, only the lowest order $\mathcal{O}(\kappa^{2})$
gravitational correction are listed, and the $\mathcal{O}(\kappa^{0})$ terms listed in Eqs.(\ref{Equ:purescalarresult-3}) are omitted. Since
we are interested only in the lowest order gravitational corrections
to the renormalization of the Higgs sector, the one-loop one particle irreducible (1PI) Feynman diagrams relevant for our concern must
satisfy three conditions. The first is all the external line particles must be Higgs field. The second condition is there must exist and
only exist a graviton in the internal lines. And the last one is that
the Feynman diagrams must be divergent.

\begin{figure}[htbp]
    \centering
    \subfigure{\includegraphics[width=1.8cm]{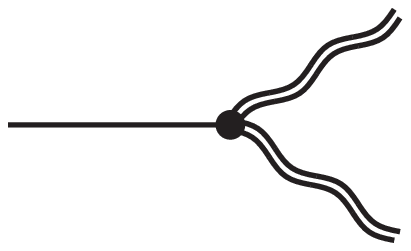}}
    \subfigure{\includegraphics[width=1.8cm]{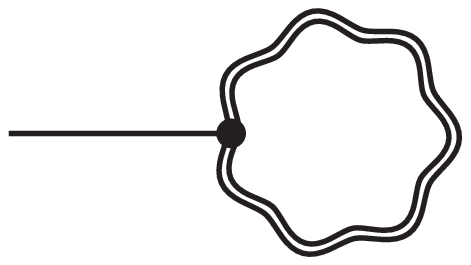}}\\
    \hspace{0.5cm}(a)\hspace{1.5cm}(b)\\
    \caption{Because of the absence of interaction vertex shown in (a),
    there is no such diagram listed in (b).}
    \label{Fig:gravitononepoint}
\end{figure}

\begin{figure}[htbp]
    \centering
    \includegraphics[width=2.5cm]{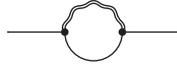}
    \caption{$\mathcal{O}(\protect\kappa^{2})$ correction
    to the two-point function. The diagram is of weight 1
    but no permutation, which leads to a factor 1.}
    \label{Fig:gravitontwopoint}
\end{figure}

\begin{figure}[htbp]
    \centering
    \subfigure{\includegraphics[width=2.5cm]{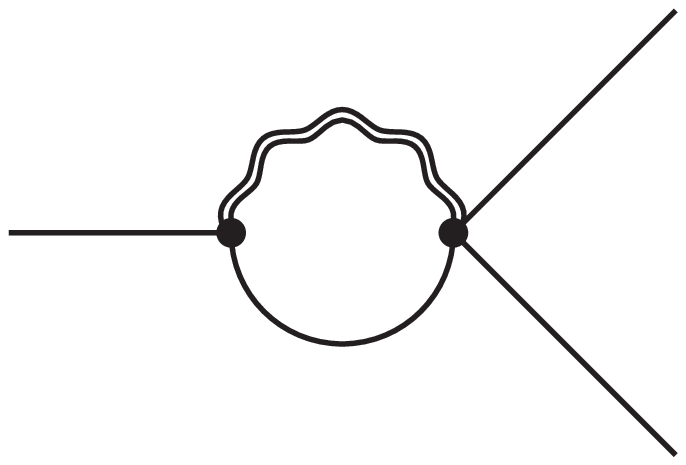}}
    \subfigure{\includegraphics[width=2.5cm]{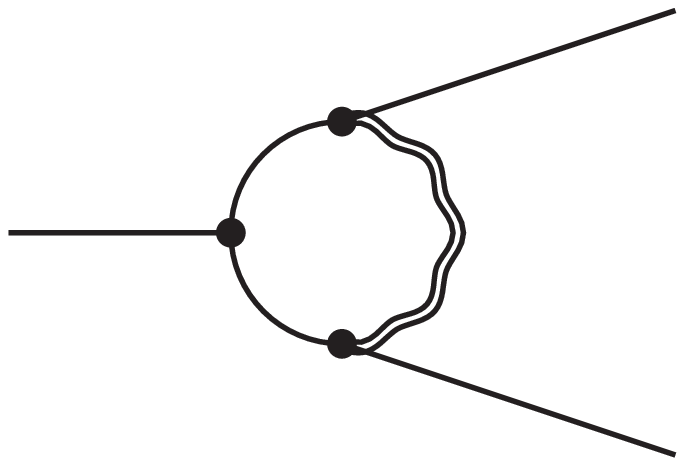}}\\
    (a)\hspace{3cm}(b)\\
    \caption{$\mathcal{O}(\kappa^{2})$ corrections to the
    three-point functions of $\phi$. (a) and (b)
    are both of weight 1 and permutation 3, which leads to a
    factor 3.}
    \label{Fig:gravitonthreepoint}
\end{figure}

\begin{figure}[htbp]
    \centering
    \subfigure{\includegraphics[width=2.5cm]{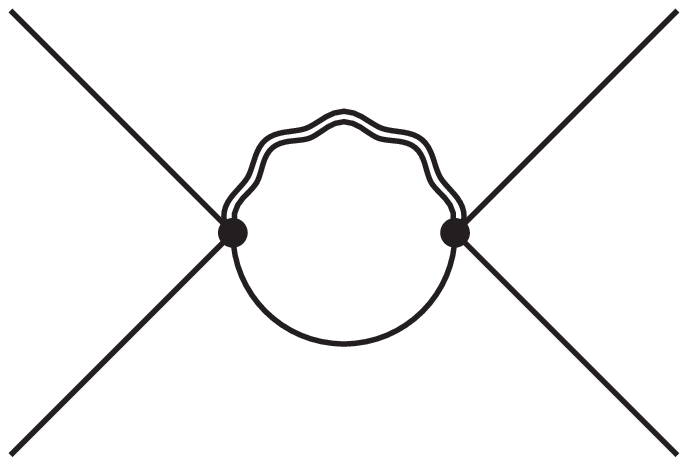}}
    \subfigure{\includegraphics[width=2.5cm]{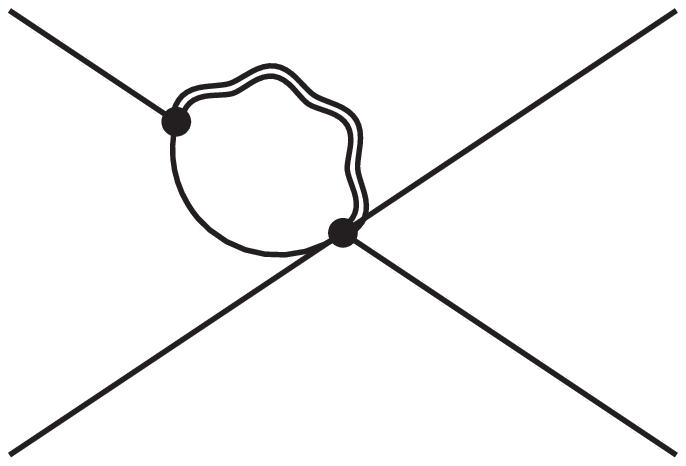}}
    \subfigure{\includegraphics[width=2.5cm]{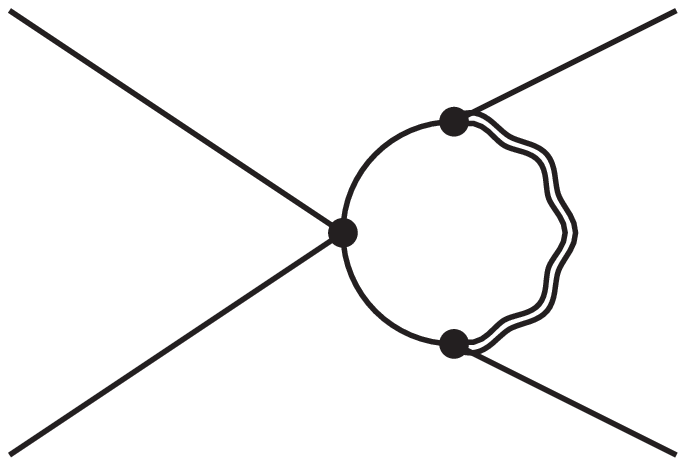}}\\
    (a)\hspace{2.5cm}(b)\hspace{2.5cm}(c)\\
    \caption{$\mathcal{O}(\protect\kappa^{2})$ corrections to
    the four-point functions of $\phi$. (a)-(c) are
    of permutations 3, 4, and 6, and all of weight 1, which
    leads to a factor of 3, 4, and 6, respectively.}
    \label{Fig:gravitonfourpoint}
\end{figure}

Because of the absence of the interaction vertex as shown in Fig.(\ref{Fig:gravitononepoint}.a) in this model, there
is no such one-loop tadpole diagram corrected by the graviton
as listed in Fig.(\ref{Fig:gravitononepoint}.b), which is
the only diagram of $\mathcal{O}(\kappa^{2})$ that contributes
to the one-point function of the Higgs field.
Hence there is no contribution from the $\mathcal{O}(\kappa^{2})$
one-loop tadpole diagram in the calculation, and then it can be
obtained that,
\begin{eqnarray}
\delta_{t}&=&0.\label{Equ:Gonepoint}
\end{eqnarray}
The diagrams for two-point, three-point and four-point
functions are listed in Figs. \ref{Fig:gravitontwopoint}-\ref{Fig:gravitonfourpoint}, respectively, where
the weight factor and permutations of external legs
have been taken into account.

As discussed above, there is no odd number of Goldstone field(s)
in every term of Lagrangian. Considering in the lowest order
gravitational correction there must exist and only exist a graviton
in the internal lines, the divergent one-loop 1PI Feynman diagram
with one graviton and one Goldstone field as internal lines and
Higgs fields as external lines is absent.

The corresponding two-point, three-point and four-point
functions are in turn,
\begin{eqnarray}
i\Gamma_{g}^{(2)}&=&
\frac{i\kappa^{2}\lambda v^{2}p^{2}}{(4\pi)^{2}\epsilon}
-\frac{i\kappa^{2}\lambda^{2} v^{4}}{(4\pi)^{2}\epsilon},
\nonumber\\
i\Gamma_{g}^{(3)}&=&
\frac{i\lambda v \kappa^{2}}{(4\pi)^{2}\epsilon}
[-3\lambda v^{2}+\frac{1}{2}(p_{1}^{2}+p_{2}^{2}+p_{3}^{2})],
\label{gravitationgreenfuctions}\\
i\Gamma_{g}^{(4)}&=&
\frac{i\lambda \kappa^{2}}{(4\pi)^{2}\epsilon}[-13\lambda v^{2}
+\frac{1}{2}(p_{1}^{2}+p_{2}^{2}+p_{3}^{2}+p_{4}^{2})].\nonumber
\end{eqnarray}

The squared-momentum-dependent terms in the $\mathcal{O}(\kappa^{2})$
gravitational corrections to the three-point and four-point
functions of the Higgs field do not contribute to the corresponding three-point and four-point counterterms and are the contributions
to the higher-derivative counterterms \cite{schuster2008dpl}, respectively. The only possible higher-derivative counterterms
for three-point and four-point functions can be derived from
the additional U(1) symmetry preserved term,
\begin{eqnarray}
\mathcal{L}_{hdct}&\sim&\sqrt{-g}~g^{\mu\nu}~
(\partial_{\mu}\Phi_{0})^{*}(\partial_{\nu}\Phi_{0}) \Phi_{0}^{*}\Phi_{0}.\label{Equ:originalhdct}
\end{eqnarray}

After the spontaneous breaking of global U(1) symmetry,
using Eqs.(\ref{Equ:metricup}), (\ref{Equ:measure}), (\ref{Equ:renormalizations}) and (\ref{Equ:relations})
and neglecting the higher order terms and terms with
gravitons, it is straightforward to get the high
derivative counterterms for Higgs field $\phi$,
\begin{eqnarray}
\mathcal{L}_{hdct}&\sim&\phi^{2}(\partial_{\mu}\phi)^{2}
+2v\phi(\partial_{\mu}\phi)^{2}
+v^{2}(\partial_{\mu}\phi)^{2}.\label{Equ:Ghdct}
\end{eqnarray}

The three terms in (\ref{Equ:Ghdct}) are related
to higher-derivative counterterms for four-point, three-point and
two-point functions, respectively. Hence part of the correction in
the two-point function is removed by the contribution of the third term
in (\ref{Equ:Ghdct}). These three terms correspond
to the following contributions to the four-point,
three-point, and two-point functions and should be added to them,
respectively:
\begin{eqnarray}
\phi^{2}(\partial_{\mu}\phi)^{2}&\sim&
-\frac{i\lambda \kappa^{2}}{(4\pi)^{2}\epsilon}
\frac{1}{2}(p_{1}^{2}+p_{2}^{2}+p_{3}^{2}+p_{4}^{2})
~=~i\delta\Gamma_{g}^{(4)},\nonumber\\
2v\phi(\partial_{\mu}\phi)^{2}&\sim&
-\frac{i\lambda v \kappa^{2}}{(4\pi)^{2}\epsilon}
\frac{1}{2}(p_{1}^{2}+p_{2}^{2}+p_{3}^{2})
~=~i\delta\Gamma_{g}^{(3)},
\label{Equ:hdctcontributions}\\
v^{2}(\partial_{\mu}\phi)^{2}&\sim&
-\frac{i\lambda v^{2} \kappa^{2}}{(4\pi)^{2}\epsilon}p^{2}
~=~-\frac{3 i\kappa^{2}\lambda v^{2}}{(4\pi)^{2}\epsilon}p^{2}
~=~i\delta\Gamma_{g}^{(2)}.\nonumber
\end{eqnarray}
A key point is that after the spontaneous breaking of global
U(1) symmetry the squared-momentum-dependent terms in the
$\mathcal{O}(\kappa^{2})$ corrections to the three-point
and four-point functions are simultaneously removed by the
higher-derivative counterterms derived from (\ref{Equ:originalhdct})
for four-point and three-point functions, respectively.

Taking into account (\ref{Equ:hdctcontributions}), from two-point
and three-point functions and Eq.(\ref{Equ:Gonepoint}), it is straightforward to get
\begin{eqnarray}
\delta_{\phi}&=&
\frac{\kappa^{2}\lambda v^{2}}{2(4\pi)^{2}\epsilon},\nonumber\\
\delta_{\lambda}&=&
-\frac{5\kappa^{2}\lambda v^{2}}{(4\pi)^{2}\epsilon},\nonumber\\
\delta_{v}&=&
\frac{2\kappa^{2}\lambda v^{2}}{(4\pi)^{2}\epsilon},
\label{Equ:gc3point}
\end{eqnarray}
while from two-point and four-point functions and
Eq.(\ref{Equ:Gonepoint}), one can obtain
\begin{eqnarray}
\delta_{\phi}&=&
\frac{\kappa^{2}\lambda v^{2}}{2(4\pi)^{2}\epsilon},\nonumber\\
\delta_{\lambda}&=&
-\frac{13\kappa^{2}\lambda v^{2}}{(4\pi)^{2}\epsilon},\nonumber\\
\delta_{v}&=&
\frac{6\kappa^{2}\lambda v^{2}}{(4\pi)^{2}\epsilon}.
\label{Equ:gc4point}
\end{eqnarray}

It is easy to find that the results for $\delta_{\lambda}$ and $\delta_{v}$ obtained from two-point and three-point functions
of the Higgs field [see Eqs.(\ref{Equ:gc3point})] and that from
two-point and four-point functions [see Eq.(\ref{Equ:gc4point})] contradict each other, which indicates that the lowest order gravitational correction to the renormalizations of the Higgs sector in this model is inconsistent.

If we reduce the complex scalar field $\Phi$ to a real scalar field (namely, $\Phi^{*}=\Phi$), the Lagrangian has a Z$_{2}$ symmetry, $\Phi\to-\Phi$, then we arrive at a real scalar $\Phi^{4}$ theory with spontaneously broken Z$_{2}$ symmetry. There is no Goldstone field after the spontaneous breaking of the discrete Z$_{2}$ symmetry. Therefore, compared with the case of the $\Phi^4$ theory with spontaneously broken global U(1) symmetry, all of the Feynman diagrams with Goldstone field $\rho$ in the pure real scalar correction part should be removed here. It is easy to check that the renormalization of the Higgs field $\phi$ self-correction is also consistent, and the result reads
\begin{eqnarray}
\delta_{t}&=&\frac{3 \lambda}{2(4\pi)^2\epsilon},\nonumber\\
\delta_{\phi}&=&0,\nonumber\\
\delta_{\lambda}&=&\frac{9 \lambda}{2(4\pi)^2\epsilon},\nonumber\\
\delta_{v}&=&-\frac{3 \lambda}{2(4\pi)^2\epsilon}.
\label{Z2purescalarresult-3}
\end{eqnarray}

For gravitational correction to $\Phi^{4}$ theory with spontaneously
broken Z$_{2}$ symmetry, the interactions between Higgs field $\phi$
and graviton $h_{\mu\nu}$, hence the related Feynman diagrams, and
the high derivative counterterms, do not change compared with the
case of $\Phi^{4}$ theory with spontaneously broken U(1) symmetry.
On the other hand, as pointed out in Sec.~\ref{corrections},
the Goldstone field does not contribute to the gravitational correction
to the Higgs sector in the case of $\Phi^{4}$ theory with spontaneously
broken U(1) symmetry. Therefore even though the Goldstone field is
absent in the case of $\Phi^{4}$ theory with spontaneously broken
Z$_{2}$ symmetry, there is no difference between these two cases for $\mathcal{O}(\kappa^{2})$ gravitational correction to the Higgs sector.
The contradiction of the gravitational correction to Higgs field still
holds.

\section{\label{comparisonanddiscussion}Comparison and Discussion}

In order to reveal the reason of the inconsistence of the gravitational
correction to the Higgs sector, we make some comparison with SQED with spontaneously broken global U(1) symmetry, whose Lagrangian takes the
form
\begin{eqnarray}
\mathcal{L}_{SQED}&=&T(\Phi,A_{\mu})-V(\Phi),
\label{SQEDlagrangian}
\end{eqnarray}
where the potential $V(\Phi)$ is given by Eq.(\ref{Equ:potential}),
and the kinetic term is
\begin{eqnarray}
T(\Phi,A_{\mu})&=&
(D_{\mu}\Phi)^{*}(D^{\mu}\Phi)-\frac{1}{4}F_{\mu\nu}F^{\mu\nu},
\end{eqnarray}
with $D_{\mu}=\partial_{\mu}+i e A_{\mu}$. The renormalizability
of this model with local U(1) transformation had been studied
in Ref.\cite{Appelquist1973}. Here we are interested in the
behavior of this model under the global U(1) transformation Eq.(\ref{Equ:transformation}). Using Eqs.(\ref{Equ:VEV}), (\ref{Equ:decomposion}), (\ref{Equ:vacuum}) and (\ref{Equ:barefield}),
and setting $\delta_{t}=0$ in Eq.(\ref{Equ:barefield}), one arrives at
\begin{eqnarray}
T(\Phi,A_{\mu})
&=&\frac{1}{2}(\partial_{\mu}\phi)^{2}
+\frac{1}{2}(\partial_{\mu}\rho)^{2}
+e^{2}v \phi A_{\mu}A^{\mu}\nonumber\\&&
+\frac{1}{2}e^{2}\phi^{2}A_{\mu}A^{\mu}
+\frac{1}{2}e^{2}\rho^{2}A_{\mu}A^{\mu}
-e \rho A^{\mu} \partial_{\mu}\phi\nonumber\\&&
+e \phi A^{\mu} \partial_{\mu}\rho
-\frac{1}{4}F^{\mu\nu}F_{\mu\nu}
+\frac{1}{2}e^{2}v^{2}A_{\mu}A^{\mu}\nonumber\\&&
+e v A^{\mu}\partial_{\mu}\rho.
\label{Equ:SQEDTterm}
\end{eqnarray}

In this case, the covariant derivative in the kinetic term $(D_{\mu}\Phi)^{*}(D^{\mu}\Phi)$ consists the gauge field
$A_{\mu}$ since the complex scalar field $\Phi$ carries charge.
This directly leads to the appearance the mixing between
$A_{\mu}$ and $\rho$ (see the last term in Eq.(\ref{Equ:SQEDTterm}))
after the spontaneous breaking of global U(1) symmetry, which is
proved to play a crucial role in the renormalization of this model.
Because of such mixing term, the gauge field will be massive after
eating the Goldstone boson and the mass of the gauge field,
which is related to the VEV, is just the reflection of the
information from the spontaneous breaking of global U(1) symmetry.
The reason that the massive gauge field can give a consistent
correction to the Higgs field $\phi$ is that the massive gauge field
contains not only the original massless gauge field but also the Goldstone field which is a part of the original complex scalar field $\Phi$.

In fact, considering the coupling constants and interactions
of three-point and four-point Higgs field self-interactions are
not independent after the spontaneous breaking of global U(1)
symmetry, one cannot expect that an ordinary alien gauge field
could give a consistent correction to such model. But only if the
parameters in the theory, e.g., mass of gauge boson $m_{A}=e v$,
is connected with the VEV of the original complex scalar field by the
mixing term between $A_{\mu}$ and $\rho$, and the interactions
among the gauge field, the Higgs field, and the Goldstone field are derived
from the spontaneous breaking of global U(1) symmetry, the
expectation for a consistent correction is reasonable.

But something changes in the case of the gravitational correction to
the Higgs sector in our model. When the complex scalar field $\Phi$
couples to gravity, the connection in the covariant derivative
of the kinetic term of the field, e.g., $\mathcal{D}_{\mu}\Phi$, vanishes, resulting in $\mathcal{D}_{\mu}\Phi\to\partial_{\mu}\Phi$.
This will directly lead to the absence of the two-point mixing
terms between the graviton and the Goldstone field. On the other hand,
considering the symmetry is spontaneously broken in the internal
charged field space, which is different from the external spacetime,
one will find this is natural that the Goldstone field can mix
with gauge field $A_{\mu}$ but not graviton $h_{\mu\nu}$. Therefore
the information of the spontaneous breaking of symmetry puts
no influence on the graviton, and the Goldstone field, which
appears after the spontaneous breaking of symmetry, no longer
contributes to the gravitational correction to the Higgs sector.
Actually this inconsistent result is not surprising when we take
into account what we considered is the correction of an alien
field, which carries no information of the original symmetry,
to the Higgs field after the spontaneous breaking of symmetry.

It is also interesting to compare a graviton in this model with
a gluon in the quantum chromodynamics. Since the Higgs field is a
color singlet, the coupling between the gluon field and the Goldstone field
is forbidden. Therefore the two-point mixing term between the gluon and
the Goldstone field is also absent. But unlike the fact that the correction
of the gluon to the Higgs sector is meaningless, gravity can couple
to Higgs field and contribute to the correction of the Higgs field
due to the fact that gravity is a reflection of the feature of
spacetime and could couple to any kind of energy.

Hence in the SM, every field that contributes to the correction
to the Higgs sector will be influenced by the spontaneous breaking of
symmetry, specifically speaking, the masses of $W^{\pm}$ and
$Z^{0}$ are related to the VEV of the original complex scalar field,
while on the other hand, the gluon and photon, whose couplings with
the Higgs field are absent, will not contribute to such a correction.
For this reason, the consistence of the gravitational correction
to the SM is questionable considering that the gravity, even not influenced by the spontaneous breaking of symmetry, will still
contribute to the correction to the Higgs sector in the SM.

\section{\label{conclusion}Conclusion}

In summary, in this paper we considered a model in which perturbatively quantized Einstein gravity couples to the $\Phi^4$ theory with spontaneously broken global U(1) symmetry and calculated the lowest order pure real scalar correction and gravitational correction to the renormalizations
of the Higgs sector. It is found that there is a contradiction in the gravitational correction to the renormalization of the Higgs sector
while the pure real scalar correction to it leads to a compatible renormalization of such model.

Based on the analysis above, the consistence for the gravitational correction to the renormalization of the Higgs sector in the SM, where the electroweak SU(2)$_{L}\times$U(1)$_{Y}$ symmetry is spontaneously broken, is open to doubt. It is expected that this contradiction will still hold there, which is the subject of future study.

\begin{acknowledgments}
We are grateful to Professor Mu-Lin Yan and Professor Dao-Neng Gao for helpful discussions. Hao-Ran Chang would like to express his special thanks to Professor Jun Yan for favorable correspondence. This work was supported by the National Natural Science Foundation of China under Grants No.11075149, No.10975128, and No.11074234.
\end{acknowledgments}

\end{CJK*}
\end{document}